**Suppression of multiferroic order in hexagonal YMn$_{1-x}$In$_x$O$_3$ ceramics**

A. Dixit[1], Andrew E. Smith[2], M.A. Subramanian[2], G. Lawes[1]

1. Department of Physics and Astronomy, Wayne State University, Detroit, MI 48201

2. Department of Chemistry, Oregon State University, Corvallis, OR 97330

**Abstract**

We have investigated the effects of substituting In for Mn on the antiferromagnetic phase transition in YMnO$_3$ using magnetic, dielectric, and specific heat measurements. We prepared a set of isostructural phase pure hexagonal YMn$_{1-x}$In$_x$O$_3$ samples having x=0 to x=0.9, which exhibit a systematic decrease of the antiferromagnetic ordering temperature with increasing In content. The multiferroic phase, which develops below T$_N$, appears to be completely suppressed for x≥0.5 in the temperature range investigated, which can be attributed solely to the dilution of magnetic interactions as the crystal structure remains hexagonal. Similar to previous reports, we find an enhancement of the magnetocapacitive coupling on dilution with non-magnetic ions.

# 1. Introduction

The hexagonal rare earth manganites, RMnO$_3$ with R=Ho, Er, Tm, Yb, Lu, or Y, have been widely studied because of their magnetoelectric properties[1-6]. YMnO$_3$, which develops ferroelectric order at T$_C$~900 K and antiferromagnetic order below T$_N$~75 K, and is therefore multiferroic below T$_N$, has been the subject of numerous investigations exploring the nature of the magnetoelectric coupling in this class of materials. Early measurements on the temperature dependent dielectric constant of YMnO$_3$ found clear anomalies associated with the antiferromagnetic transition, pointing to coupling between the ferroelectric and magnetic orders [1]. Optical studies on YMnO$_3$ also find clear evidence for coupling between the ferroelectric and antiferromagnetic domain walls [3], with measurements on the related compound HoMnO$_3$ demonstrating control of the magnetic phase using a static electric field [7]. The microscopic origin of the spin-charge coupling in YMnO$_3$ has also been investigated in detail using Raman scattering [8-9] and magnetoelastic measurements [6]. Raman studies on YMnO$_3$ thin films provide evidence for spin-phonon coupling below T$_N$ [9], while neutron measurements of spin excitations highlight the importance of magnetoelastic coupling in this system [6]. Theoretical studies on YMnO$_3$ point to the importance of the displacement of the Y$^{3+}$ ions [10], which is facilitated by hybridization of the Y 4d orbitals with the O 2p orbitals and the buckling of the MnO$_5$ bipyramids [11].

Because the multiferroic properties of YMnO$_3$ are relatively well understood, it is illuminating to determine how the magnetic and ferroelectric characteristics change on substituting on the both the Y and Mn sites. The ferroelectric distortion and magnetic ordering are found to be suppressed when Zr replaces Y [12], which also decreases the carrier concentration [13]. The carrier concentration can be increased by replacing Y with Mg or Li, the latter of which also produces weak ferromagnetic behaviour [13]. The magnetocapacitive coupling in $YMnO_3$ is found to increase when Mn is replaced by modest amounts of Ti [14] or Ga [15], although increasing the Ti content above approximately 20% is found to suppress the ferroelectric P63cm structure in favor of a centrosymmetric R$\underline{3}$c structure [14]. Previous studies have shown that Ga doping, which does not produce any change in crystal structure, significantly affects the magnetocapacitive coupling along the c-axis, while producing much more modest changes for the dielectric constant measured perpendicular to the c-axis [15]. Substituting non-magnetic Ga on the $Mn^{3+}$ sites slightly increases the Curie temperature while systematically reducing the antiferromagnetic ordering temperature [16]. However, the reduction of the spin-lattice interactions in the Ga doped system increases the thermal conductivity, despite the introduction of additional impurity scattering [17].

In order to more fully investigate the magnetoelectric properties of $YMnO_3$ at the antiferromagnetic phase transition with substitution on the Mn site we investigated solid solutions of $YMnO_3$ and $YInO_3$. Both of these compounds can be stabilized in the hexagonal structure, having similar M-O bond lengths in the basal plane but with the Mn-O apical bond length being considerably shorter than the In-O apical bond length [18]. Because both end members of the $YMn_{1-x}In_xO_3$ series can be prepared in the hexagonal structure, we presume that the structural instabilities that affect other systems, such as $YMn_{1-x}Ti_xO_3$, can be avoided over the entire range of compositions 0≤x≤1. Furthermore, because it is possible to stabilize high concentrations of In in this structure, the antiferromagnetic transition can be suppressed completely for large x, which allows us to probe the properties of the paramagnetic system down to low temperatures. Furthermore, there is increased interest in In substituted $YMnO_3$ specifically, as this system has also been identified as an inexpensive, minimally toxic blue pigment [18].

## 2. Experimental

Ceramic samples of $YMn_{1-x}In_xO_3$ for compositions ranging from x=0 to 1 were prepared by heating a stoichiometric ratio of $Y_2O_3$, $Mn_2O_3$, and $In_2O_3$ at elevated temperatures with intermediate grindings. Full details of the synthesis procedure are available elsewhere [18]. X-ray powder diffraction data (XRD) was collected at room temperature with a RIGAKU MINIFLEX II diffractometer using Cu Kα and a graphite monochromator. The XRD patters are shown in Fig. 1a. All peaks can be fully indexed to the hexagonal P63cm structure, with no evidence for the formation of any secondary phases. A number of peak positions change systematically with the Mn:In ratio and the variation is illustrated in a Vegard's law plots in Figs 1b and 1c. These show the expected behavior as $a_{[YMn1-xInxO3]}=a_{[YMnO3]}(1-x)+a_{[InMnO3]}x$. The linear variations in *a* and *c* lattice parameter are consistent with the length of the apical M-O bond increasing as the In fraction of the compound increases.

The central focus of this investigation was to determine how the magnetic, thermodynamic, and dielectric properties of $YMn_{1-x}In_xO_3$ at the antiferromagnetic phase transition depend on x over a wide range of compositions. We used a Quantum Design MPMS SQUID magnetometer and a

Quantum Design Physical Property Measurement System (PPMS) to measure the magnetic properties of approximately 30 mg of the powder samples. For the specific heat measurements, we thoroughly mixed approximately 50 mg of each sample with the same mass of Ag powder then pressed this composite into small disks to ensure good thermal contact throughout the entire sample. The heat capacity was then measured using the standard option on the Quantum Design PPMS. This system was also used to provide temperature and magnetic field control for the capacitance measurements, which were done on cold-pressed pellets prepared from the YMn$_{1-x}$In$_x$O$_3$ powders having electrodes fashioned from silver epoxy. These dielectric measurements were done at a measuring frequency of 30 kHz. Separate measurements (not shown) indicated that the frequency response of these samples was flat in the region between 1 kHz and 100 kHz.

## 3. Results and discussion

The temperature dependent magnetic susceptibilities for the x=0, x=0.25, x=0.5, and x=0.75 samples measured at $\mu_0H$=0.5 T are plotted in Fig. 2a. There is a systematic decrease in the high temperature susceptibility with increasing x, consistent with replacing Mn$^{3+}$ with non magnetic In$^{3+}$ ions. The Curie-Weiss temperature, $\Theta_{CW}$, estimated from the high temperature susceptibility varies systematically from $\Theta_{CW}$=-337 K for x=0 to -325 K, -240 K, and -83 K for x=0.25, x=0.5, and x=0.75 respectively. The effective moment per Mn estimated from the high temperature susceptibility is 4.2±0.5 $\mu_B$ for all samples. At low temperatures, the paramagnetic Curie-Weiss behaviour for the x=0.5 and x=07.5 samples leads to a larger susceptibility than in the magnetically ordered x=0 and x=0.25 samples. The magnetic anomalies associated with the development of antiferromagnetic order for the x=0 and x=0.25 samples can be seen more clearly in the inset to Fig. 3a, which plots the susceptibility corrected for the background paramagnetic behavior. The magnetic transition occurs at $T_N$=72 K for $YMnO_3$, consistent with previously observed values [1]. The transition temperature is suppressed as Mn is replaced by non-magnetic In, falling to $T_N$=42 K for the $YMn_{0.75}In_{0.25}O_3$ sample. The magnitude of this suppression is what has been observed previously in Ga substituted $YMnO_3$ having a similar fraction of non-magnetic Ga replacing the $Mn^{3+}$ ions [15]. We do not observe any clear anomalies in the magnetization curves for samples having x≥0.5, suggesting that the antiferromagnetic transition temperature is pushed to below T=20 K, where the paramagnetic Curie tail obscures the magnetization data.

We plot the temperature dependent specific heat for different $YMn_{1-x}In_xO_3$ samples in Fig. 2b, all measured at zero applied field. This plot shows only the contribution from the magnanite samples; the background silver contributed was measured separately and removed. The samples having smaller In fractions exhibit a clear peak in heat capacity at the antiferromagnetic phase transition, while the In-rich samples do not show any clear anomalies. For x=0.5 and above, the heat capacity is roughly the same for all compositions, confirming that the differences in specific heat arise solely from the magnetic order of the Mn ions and not from any significant difference in lattice contributions. Estimating the lattice contribution to the $YMnO_3$ heat capacity from the $YMn_{0.5}In_{0.5}O_3$ data, and neglecting the slight difference in mass between Mn and In, we find that the magnetic entropy between 25 K and 85 K associated with Mn spin ordering is approximately 5.3 J/mole K. This is considerably smaller than the Rln5=13.4 J/mole K expected for the full ordering of S=2 $Mn^{3+}$ ions, which has been observed in previous studies of $YMnO_3$ samples [4]. We attribute this discrepancy to the considerable entropy developing above T=85 K, which is excluded from our estimate, as well as errors associated with correcting for the silver background contributions.

One of the signatures of the coupling between the magnetic and ferroelectric order parameters in $YMnO_3$ is the presence of a dielectric anomaly at the antiferromagnetic transition temperature. The relative temperature dependent dielectric constants for selected $YMn_{1-x}In_xO_3$ samples, scaled to the value at T=80 K just above the transition temperature for $YMnO_3$, are plotted in Fig. 3a. The slight anomalies in the dielectric constant coincident with antiferromagnetic ordering, visible mainly as a small change in slope, are indicated by arrows. The value of $T_N$ is slightly reduced for $YMn_{0.9}In_{0.1}O_3$ as compared to $YMnO_3$, and still further reduced in $YMn_{0.75}In_{0.25}O_3$. There is no dielectric anomaly in the temperature range investigated for $x \geq 0.5$, consistent with our magnetic and specific heat measurements, neither of which show any phase transition. The dielectric constant for the x=0.25 sample varies more with temperature than the other samples. We tentatively attribute this to the fact that the dielectric loss (not shown) for this particular sample is slightly larger than the loss for other compositions, rather than reflecting any intrinsic response to In substitution.

Previous studies on $YMnO_3$ doped with Ga found evidence for a significant enhancement of the magnetocapacitance, which was attributed, at least in part, to the non-magnetic ions relieving the frustration leading to larger magnetoelectric coupling along the c axis [15]. To explore whether similar behavior would develop in In doped $YMnO_3$ and whether the enhancement in the coupling would persist to larger fractions of non-magnetic ions, we measured the temperature and magnetic field dependence of the capacitance for the different samples. The field dependence of the relative change in capacitance measured at T=10 K is shown for the different samples in Fig. 3b. This temperature is below the magnetic ordering temperature for the x=0, x=0.1, and x=0.25 samples, and above any possible ordering temperature for the x=0.75 sample. Because of the very capacitive signal in these samples, the data were rather noisy, so the results shown in Fig. 3b plot the results after smoothing. The solid bars on the x=0, x=0.5, x=0.75 curves provide a visual estimate of the spread in the data before smoothing. We find that the x=0 and x=0.1 samples have negligible magnetocapacitive coupling, with some small shift in the capacitance observed for the x=0.25 sample. The x=0.5 and x=0.75 samples show significant larger field-induced shifts in capacitance, with the change in capacitance reaching over 0.05% in an applied field of $\mu_0 H$=8 T. We emphasize that these samples having a large In fraction do not order magnetically, so this large magnetocapacitance represents a coupling in to paramagnetic spins. These changes may be consistent with those observed for single crystal $YMn_{0.7}Ga_{0.3}O_3$ samples, where the shift reached approximately -0.25% for E || c and 0.025% for E ⊥ c at similar magnetic fields [15]. As the the magnetocapacitive shift in Fig. 4b represents a powder average of the response for E || c and E ⊥ c, we would expect the magnitude of the shift to be intermediate between the values for the single crystal Ga:$YMnO_3$ sample.

One of the most striking features of the magnetocapacitive shift shown in Fig. 3b is the change in sign between different samples. At a mean field level, the relative magnetocapacitive shift is expected to be proportional to the square of the applied magnetic field, $\Delta C/C \approx \gamma H^2$, arising from the lowest order term coupling P and H allowed by symmetry in the free energy [19]. A fit to this functional form for the x=0.75 sample (allowing a small offset) is indicated by the dashed line in Fig. 3b. We fit the temperature dependence of the coupling parameter gamma for the

x=0.5 and x=0.75 samples, selected as these show the largest magnetocapacitive shifts, having roughly similar magnitudes, and remain paramagnetic over the entire temperature range, and plot the results in Fig. 3c. We find that $\gamma(T)$ decreases systematically with increasing temperature for the x=0.5 sample, which we attribute to first approximation to the reduction in the magnetic susceptibility. The response for the x=0.75 sample however shows a rather different temperature dependence, with a negative magnetocapaticive shift at low temperatures crossing over to a positive shift at higher temperatures. Repeated measurements show qualitatively similar behaviour for the temperature dependence of the coupling, although the magnitude of $\gamma(T)$ shows some small changes.

It has been suggested that since both the dielectric response of layered $YMnO_3$ is highly anisotropic, the magnetocapacitive coupling may be different parallel to and perpendicular to the c axis, leading to magnetocapacitive shifts as $\Delta C_\parallel/C_\parallel \approx \gamma_\parallel H^2$ and $\Delta C_\perp/C_\perp \approx \gamma_\perp H^2$ (neglecting contributions from the sublattice magnetization, which are expected to be absent for our paramagnetic x=0.5 and x=0.75 samples) [15]. Studies on $YMn_{1-y}Ga_yO_3$ find that the $\gamma_\parallel$ component measured at T=5 K changes sign as y changes from 0 to 0.3, leading to both positive and negative magnetocapacitive shifts in the compound series [15]. As the capacitance plotted in Fig. 3b represents the powder average of the anisotropic response, a change in the sign of $\gamma(T)$ could be produced by a change of sign in $\gamma_\parallel(T)$. However, since both the x=0.5 and x=0.75 samples are expected to be paramagnetic over the entire temperature range considered, it is unclear why there would be any change in sign, either as a function of In fraction or as a function of temperature. Understanding the mechanisms responsible for the change in sign of the low temperature magnetocapacitive coupling between the x=0.5 and x=0.75 samples may help to clarify the nature of the spin-charge coupling in the parent $YMnO_3$ compound.

We summarize the results of our magnetic, dielectric, and specific heat studies in Fig. 4, which plots the antiferromagnetic ordering temperature estimated from these different measurements against the In content of the sample. The transition temperatures extracted using these different measurement techniques all agree, confirming that there is a single magnetic transition giving rise to the dielectric anomalies even in the In substituted samples. The magnitude of the suppression of $T_N$ in $YMnO_3$ with In substitution is very similar to what was observed previously with Ga substitution, where the antiferromagnetic order temperature is suppressed to $T_N$=35 K with 30% Ga substitution [15]. This scaling of $T_N$ with x can be understood in terms of simple site dilution by non-magnetic ions. The initial slope of the suppression of $T_N$ with In substitution, $p = -(dT_N/dx)/T_N$, is approximately 1.6. This reduction in $T_N$ on substituting non-magnetic ions is larger than one would typically expect for a three dimensional system, where $p \approx 1.2$ [20], and is more consistent with that expected for Ising spins in two dimensions, where p normally ranges from 1.6 to 1.9 [21]. This confirms the importance of the in-plane interactions among $Mn^{3+}$ ions in developing long range antiferromagnetic order.

## 4. Conclusion

In conclusion, we have characterized the magnetic, dielectric, and thermodynamic properties of In substituted $YMnO_3$ to investigate the suppression of the multiferroic phase developing below $T_N$. We find that the magnetodielectric coupling associated with the multiferroic phase persists in the In substituted samples, but that $T_N$ decreases approximately linearly with increasing In fraction. The antiferromagnetically ordered state in $YMn_{1-x}In_xO_3$ appears to vanish for $x \geq 0.5$ and the behaviour of the initial suppression is consistent with two dimensional spin systems. There is a significant enhancement in the magnitude magnetocapacitive coupling with In substitution, which persists in the paramagnetic compounds, although the sign of this coupling varies with both the In fraction x and temperature through some mechanism that is not at all clear at this time. Taken collectively, these measurements confirm that replacing Mn by non-magnetic In simply suppresses magnetic ordering without introducing any new phase transitions while simultaneously enhancing the magnetocapacitive coupling. Since it is possible to form a complete solid solution series of $YMn_{1-x}In_xO_3$ compounds, more detailed measurements on the microscopic properties of this system may help to clarify the mechanisms giving rise to the multiferroic behaviour in the parent compound.

**Acknowledgements**

GL would like to acknowledge support from NSF through DMR-06044823. MAS would like to acknowledge support from NSF through DMR-0804167 and Air Force Research laboratory (FA8650-05-1-5041). AES is supported by NSF-IGERT grant. We would also like to acknowledge helpful conversations with Arthur Sleight and Nicola Spaldin.

**Figure captions**

**Figure 1**. (a) X-ray diffraction spectra for the $YMn_{1-x}In_xO_3$ samples, as indicated by the legend on the right of the figure. (b) and (c) Vegard's law plots for $YMn_{1-x}In_xO_3$ showing the systematic change in lattice parameter with In content, as labeled.

**Figure 2**. (a) Temperature dependence of the magnetic susceptibility measured at $\mu_0H=0.5$ T for different $YMn_{1-x}In_xO_3$ samples, as indicated. The inset plots the susceptibility less a background correction to more clearly emphasize the magnetic anomalies in the x=0 and x=0.25 samples. (b) Specific heat, plotted as C/T versus T, for the $YMn_{1-x}In_xO_3$ samples, as indicated.

**Figure 3.** (a) Relative change in capacitance for $YMn_{1-x}In_xO_3$ samples as labeled. The dielectric anomalies associated with the magnetic ordering transitions are indicated by the dashed arrow. (b) Magnetic field dependence of the capacitive response for different samples, as indicated. The solid bars on the x=0, x=0.5, and x=0.75 curves provide a visual estimate of the scatter in the data before smoothing. The dashed line on the x=0.75 curve shows the fit described in the text. (c) Temperature dependence of the magnetocapacitive coupling constant (as described in the text) for the x=0.5 and x=0.75 $YMn_{1-x}In_xO_3$ samples.

**Figure 4** Temperature dependence of the antiferromagnetic ordering transition on x in $YMn_{1-x}In_xO_3$. The different symbols represent the transition temperature extracted from different measurements, as indicated. The dashed line shows the curve used to estimate the initial suppression of $T_N$, as described in the text.

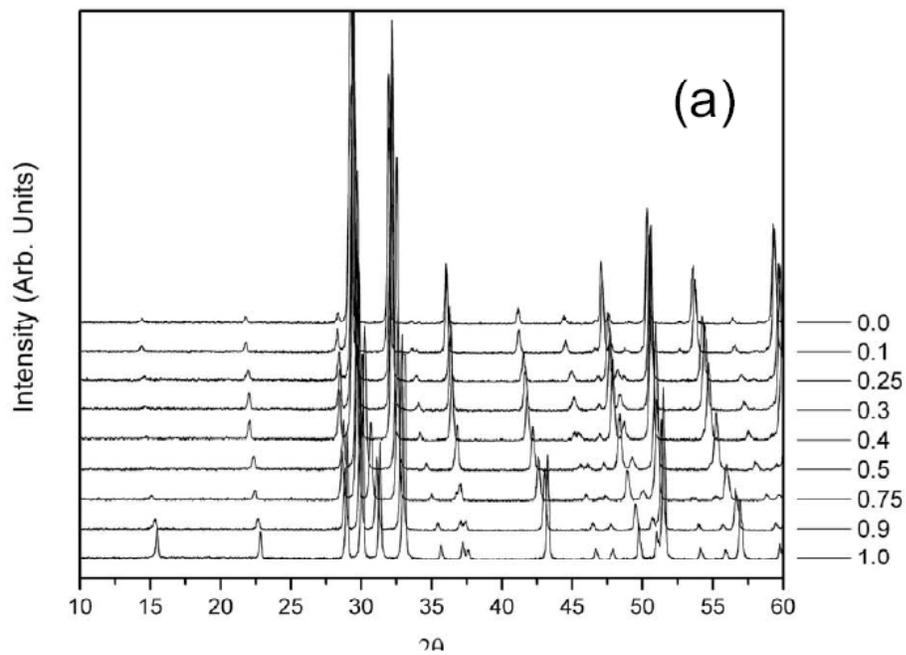
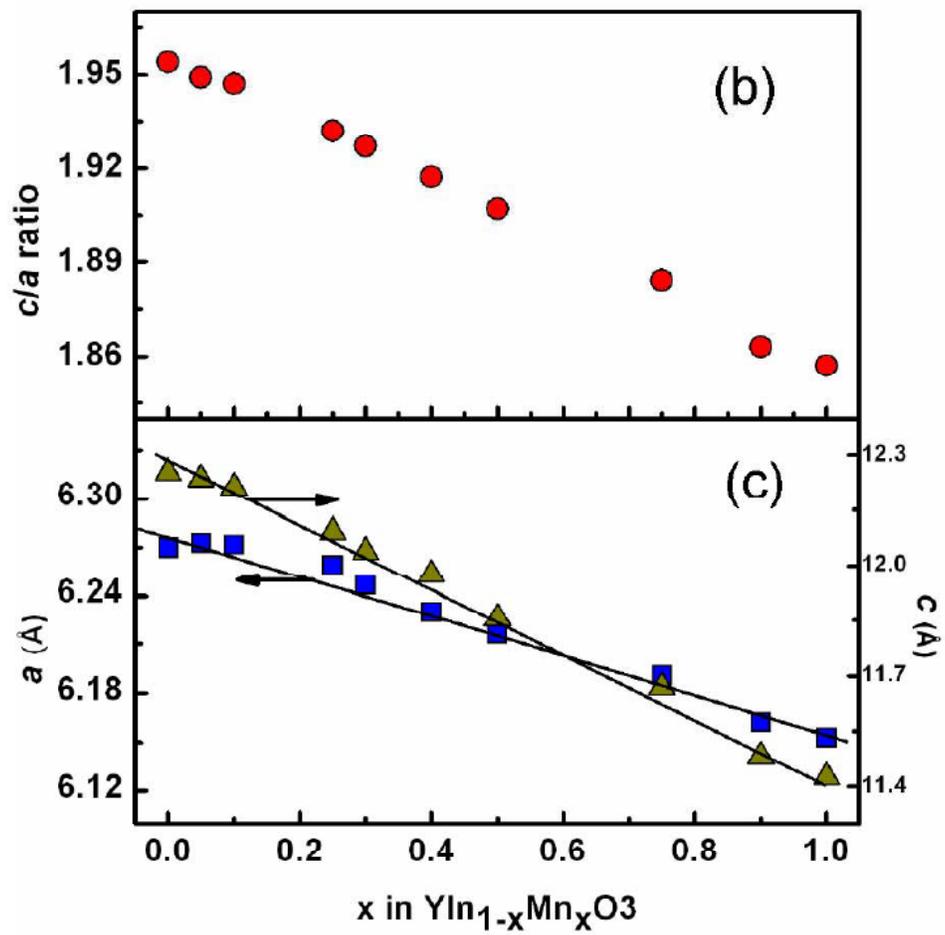

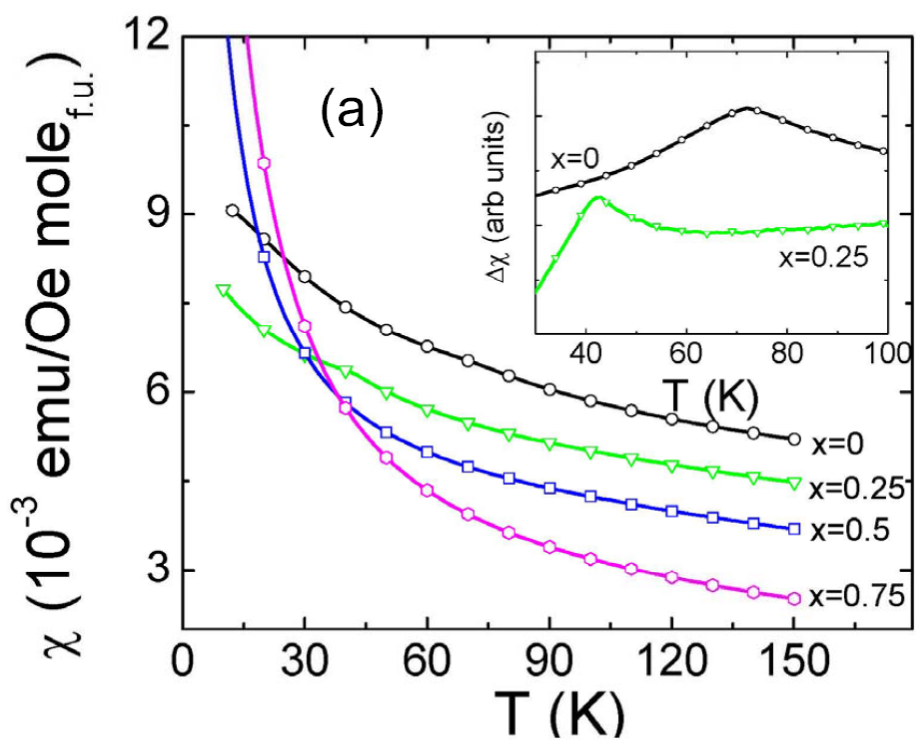
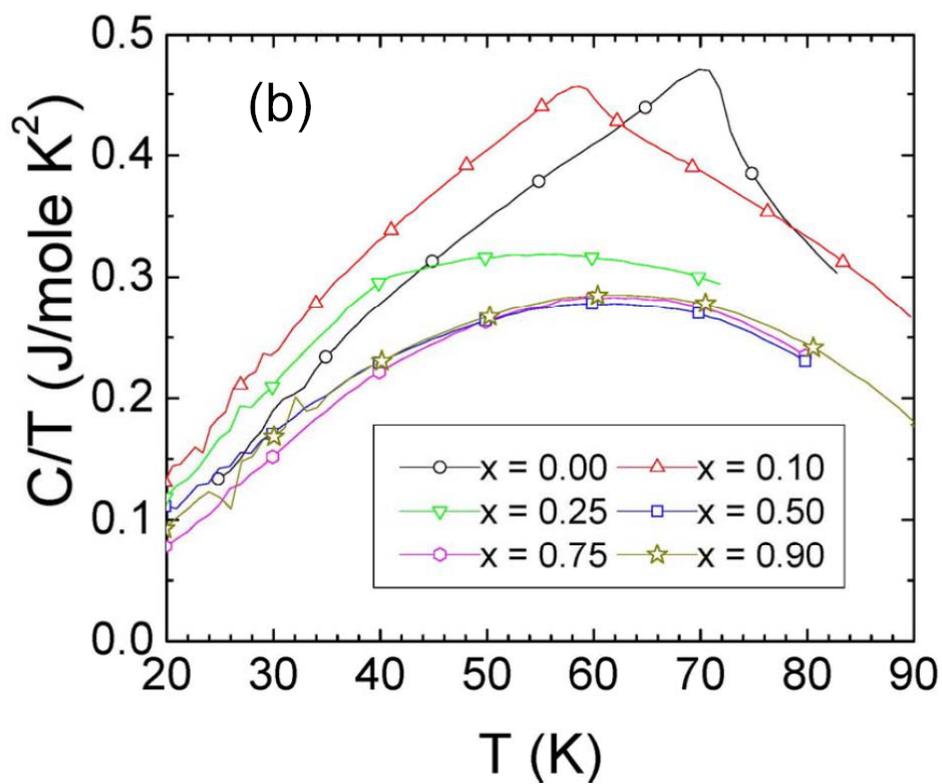

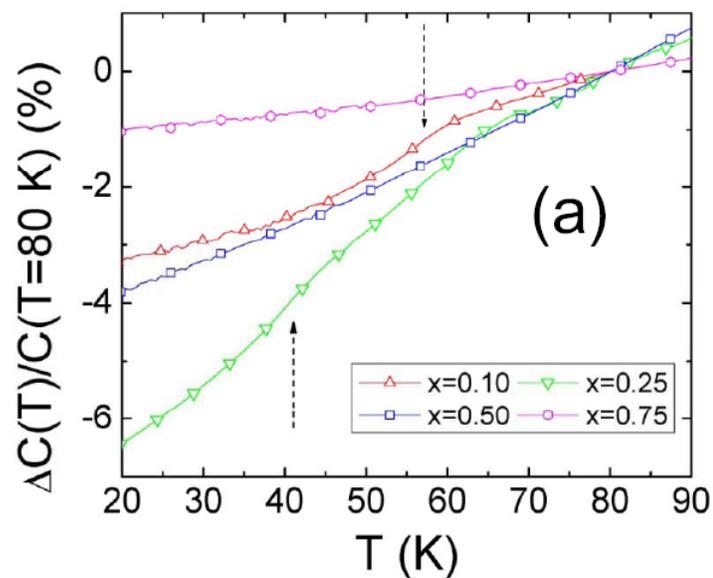
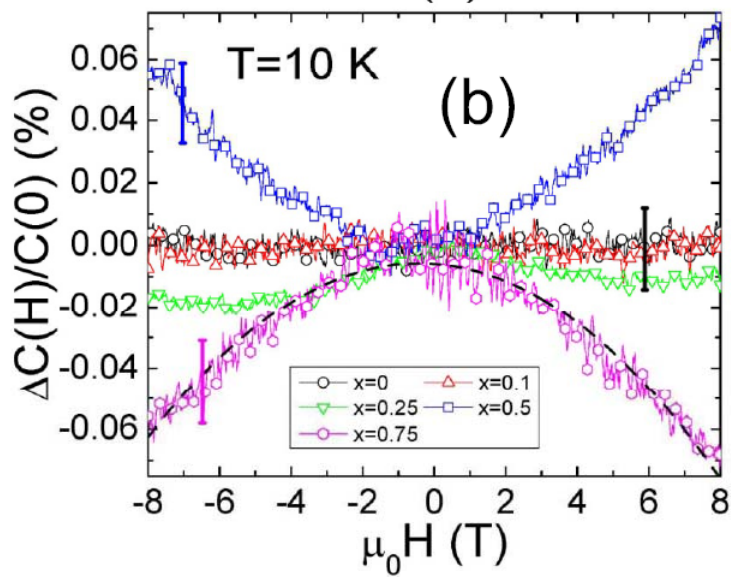
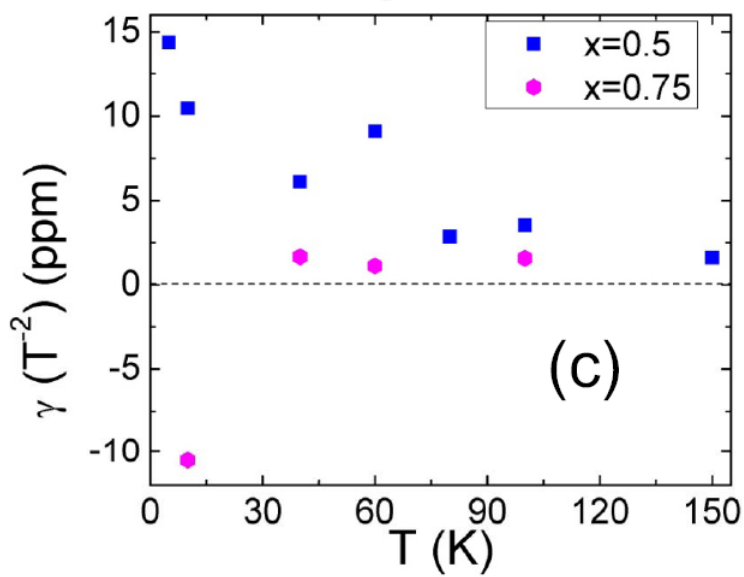

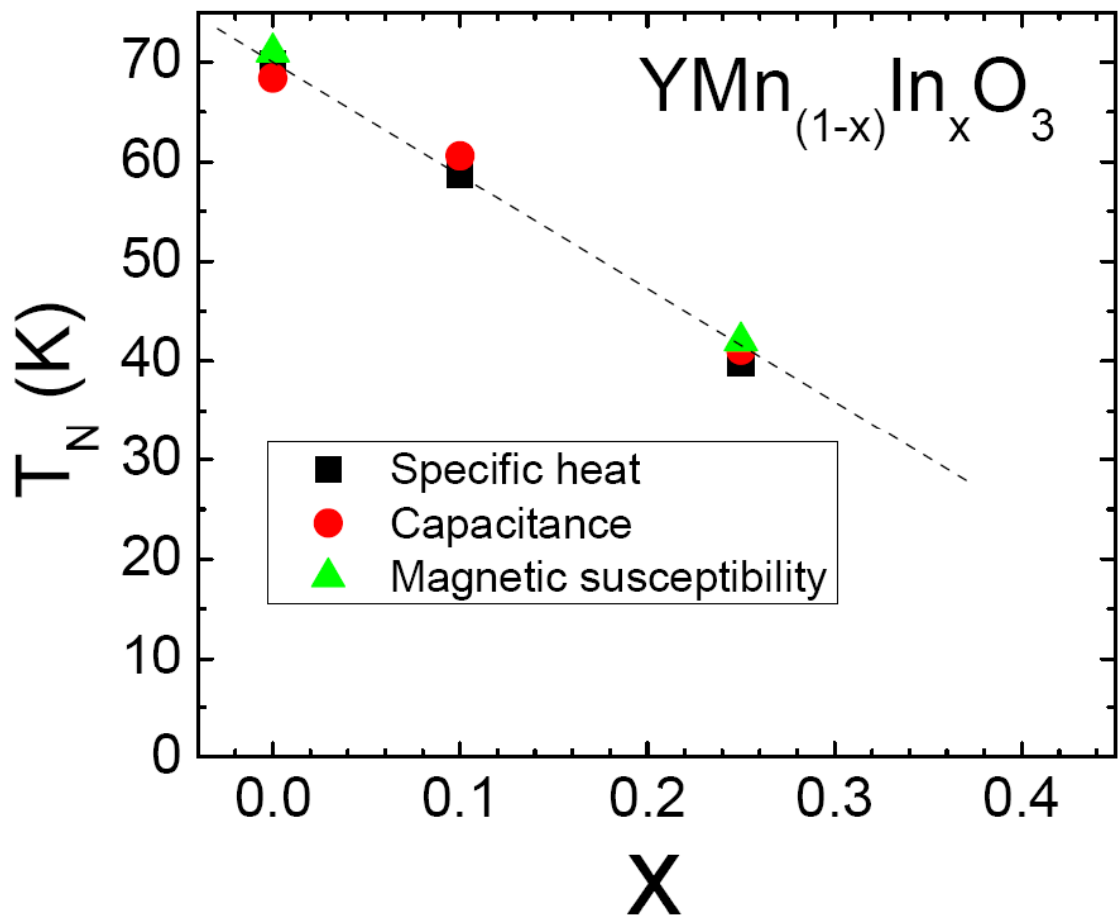